\documentclass[sigchi]{acmart}

\AtBeginDocument{%
  }

\setcopyright{acmlicensed}
\copyrightyear{2025}
\acmYear{2025}
\acmDOI{XXXXXXX.XXXXXXX}
\acmConference[Purposeful XR: Affordances, Challenges, and Speculations for an Ethical Future at CHI '25]{April 26,
  2025}{Yokohama, Japan}
\begin{document}

\title{Bridging Generations: Augmented Reality for Japanese Wartime Oral History}

%%
%% The "author" command and its associated commands are used to defines
\author{Karen Abe}
\affiliation{%
  \institution{Independent, Tokyo}\country{Japan}
}
\email{prod.karenabe@gmail.com}

\renewcommand{\shortauthors}{Karen Abe}

\begin{abstract}
In this position paper, the author presents a process artifact that aims to serve as an archival and educational tool that revitalizes World War II oral histories in Japan. First, the author introduces the historical background and how the work is informed by the positionality of the author. Then, the author presents features of the artifact using references to interview footage of the author’s grandmother and grandaunt sharing their firsthand accounts of the 1945 Tokyo Air Raids. The affordances and barriers of this application of augmented reality is discussed and a included is a list of questions to be posed at the workshop.
\end{abstract}

\begin{teaserfigure}
  \includegraphics[width=\textwidth]{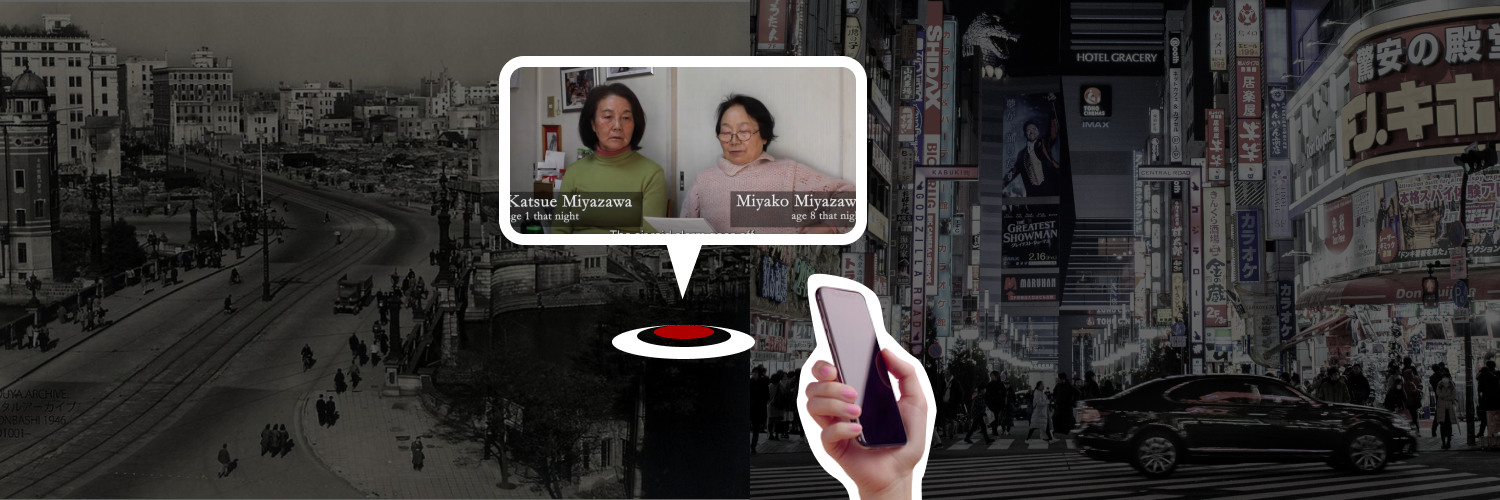}
  \caption{Mobile augmented reality showcasing interview footage over split image of 1946 Tokyo (left) and 2019 Tokyo (right).}
  \Description{Street view of Tokyo, half photograph and half draw. An image of a hand with a mobile device and image overlay pop-up of an interview of two eldery women.}
  \label{fig:teaser}
\end{teaserfigure}

\received{20 February 2007}
\received[revised]{12 March 2009}
\received[accepted]{5 June 2009}

\maketitle

\section{Introduction}
Japan faces a dual crisis: an aging population and a declining birth rate \cite{jack_issue_2016}. Although this phenomenon means Japan faces a plethora of social and political issues, another reality is that the stories of World War II are at risk of fading away. With one of the highest life expectancies in the world—nearly 84 years \cite{world_bank_life_2022}—many who lived through the war are now in their final years. At the same time, the younger population is shrinking, with projections indicating that by 2060, over 40\% of Japan’s population will be aged 65 or older\cite{ndl_birth_2004}. This demographic shift means fewer opportunities for firsthand accounts to be passed down, creating an urgent need to preserve these narratives before they are lost.

This urgency is amplified by shifting cultural and political landscapes—such as ongoing debates over the revision or reinterpretation of the Japans constitution's Article 9. Article 9, enacted post-World War II, renounces war and prohibits the maintenance of military forces, stating that "land, sea, and air forces, as well as other war potential, will never be maintained”\cite{akiyama_redefining_2014}. In 2014, the Japanese Cabinet reinterpreted Article 9 to allow for collective self-defense force. As Japan considers to increase its military presence while many elders who lived the war are against this measure — it has become increasingly important to preserve firsthand war testimonies to ensure history remains accessible and relevant to future generations.

Therefore, augmented reality can serve as an archival and educational tool that revitalizes World War II oral histories, allowing younger generations to engage with firsthand wartime narratives in immersive, interactive ways that bridge the digital and physical worlds.

Reflecting on the positionality that informs the creation of this work, the author is a Japanese-American independent media artist with direct familial ties to Japan's wartime history. From a young age, the author’s grandmother and grandaunt have passed down firsthand accounts of their experience surviving the Tokyo Air Raids during World War II. However, in recent years, as both elders have been affected by dementia — a common symptom in the Japanese aging population, accessing these memories and clear communication has grown to be difficult. Given these circumstances, the author feels personal urgency to support the preservation of Japan's oral histories.

\section{Related Work}
\textit{Fukushima Now} is a virtual reality (VR) experience that focuses on the aftermath of the 2011 Fukushima, Japan's nuclear disaster, providing an immersive environment that allows users to explore the ongoing impacts of the disaster on both the environment and the community\cite{khan_fukushima_2024}.

One of the most striking aspects of \textit{Fukushima Now} is its ability to evoke emotions through immersive storytelling. As noted by the director of the project “lot of moments in the experience where the locals have trouble remembering the past. And it's a natural case of as you go farther and farther away from a moment… there’s no history for it, except in the minds and memories of the people there”\cite{khan_fukushima_2024}. This quote speaks directly to the value of oral history and how projects like \textit{Fukushima Now} allows these personal histories to take center stage. Ultimately, the true history of Fukushima is carried not by textbooks or statistics, but by the lived experiences of the people who were affected.

\textit{Fukushima Now}’s approach underscores the importance of how mixed reality experiences can shape our understanding of history and bridge the past and present. Similarly, this project follows \textit{Fukushima Now}'s approach to a guided memory-driven experience and leverages augmented reality to embed oral histories with contemporary physical locations.

\section{Process Artifact}
\subsection{Features}
The prototype will present an augmented reality (AR) storytelling experience that uses video and audio footage of elders sharing their firsthand accounts of wartime experiences. Specifically, the prototype will feature interview footage with the author’s grandmother Katsue and grandaunt Miyako \cite{abe_doc_2020}. These interviews will be integrated with public Visual Positioning System (VPS) locations to link oral history snippets to contemporary sites. For example, in the interview, grandmother Katsue and grandaunt Miyako recount how the Kanda River (Kandagawa) was “filled with dead bodies” \cite{abe_doc_2020} during the Tokyo Air Raids. This powerful memory will be geo-located to the current position of the river, enabling users to connect the past to the present. For non-geolocated stories, specific objects or visuals will trigger AR interactions. For instance, a water fountain could activate a narrative about water scarcity during the war.

\subsection{AR Affordances and Their Impact}
Traditional oral history archives, such as museums and documentaries, often require physical presence or limited institutional access. AR expands accessibility by allowing users to engage with historical narratives through their personal devices. The interactive nature of AR fosters deeper engagement  as it offers users the autonomy to explore stories at their own pace. This feature may be suited for younger generations who seek to learn history outside of a traditional classroom context.

\subsection{\textbf{Technological and Future Advances Needed}}
For wider adoption, the experience should be compatible across various devices, ensuring accessibility beyond high-end hardware. Additionally, to ensure the geo-locative feature does not become a barrier for learning, scalability of the project would be critical in being able to apply this experience in a smaller classroom or educational settings.

\subsection{\textbf{\textbf{Potential Harms and Ethical Considerations}}}
Building strong relationships with interviewees is essential to ensuring that personal narratives are shared with care and consent. Privacy must be carefully considered, along with the need to collect diverse experiences to provide a holistic and inclusive understanding of wartime history. As seen in the \textit{Fukushima Now} project, fostering trust and rapport can help address potential skepticism or hesitation toward a technology-driven format \cite{khan_fukushima_2024}. By prioritizing these personal connections, the project can create a space where interviewees feel comfortable sharing their stories in an authentic and meaningful way.

\section{Discussion Questions}
\begin{itemize}
\item In an era where Japanese teachers struggle to adapt to digital teaching methods \cite{yoshioka_digitalization_2024}, how do we ensure mixed reality tools are well integrated in the curriculum and can further reach the younger generation in Japan?
\item How can projects like this ensure consent, accuracy, and cultural sensitivity in presenting oral histories?
\item What onboarding strategies can help engage interviewees, collect stories, and ensure the message is impactful?

\end{itemize}
\begin{verbatim}
\end{verbatim}

  \bibliographystyle{ACM-Reference-Format}
  \bibliography{CHI-AR-25}

\end{document}